\documentclass[referee,sn-basic]{sn-jnl}

\usepackage{graphicx}%
\usepackage{multirow}%
\usepackage{amsmath,amssymb,amsfonts}%
\usepackage{amsthm}%
\usepackage{mathrsfs}%
\usepackage[title]{appendix}%
\usepackage{xcolor}%
\usepackage{textcomp}%
\usepackage{manyfoot}%
\usepackage{booktabs}%
\usepackage{algorithm}%
\usepackage{algorithmicx}%
\usepackage{algpseudocode}%
\usepackage{listings}%

\begin{document}

\title[Article Title]{Reverberation responses in light curves of the SBS1520+530 quasar }

\author*[1]{\fnm{Liudmyla} \sur{Berdina}}\email{lberdina@rian.kharkov.ua}

\author[1,2]{\fnm{Victoria} \sur{Tsvetkova}}\email{tsvetkova@astron.kharkov.ua}
\equalcont{These authors contributed equally to this work.}

\affil*[1]{ \orgname{Institute of Radio Astronomy of the National Academy of Sciences of Ukraine}, \orgaddress{\street{4 Mystetstv}, \city{Kharkiv}, \postcode{61002}, \country{Ukraine}}}
\affil[2]{\orgname{Institute of Astronomy of Kharkiv National University}, \orgaddress{\street{Sumskaya 35}, \city{Kharkiv}, \postcode{61022}, \country{Ukraine}}}

\abstract{ \textbf{Purpose:} We aim to investigate an accretion disk (AD) temperature profile of a doubly lensed quasar SBS 1520+530. Temperature profiles of accretion disks are known to be diagnostic of many of the physical properties of AGNs and quasars.

\textbf{Methods:} Our approach involves application of the photometric reverberation mapping to the light curves of SBS 1520+530 obtained in the Johnson-Cousins $V$ and $R$ filters. The RM method implies that the time shift between the light curves taken in different spectral ranges determines the light travel time between the AD zones with different physical conditions. In determining the time shifts, we applied a method based on some useful properties of the orthogonal polynomials.

\textbf{Results:} The variations in filter $R$ lag those in filter  $V$, with 1.25$\pm$0.63 days for the inter-band time shift averaged between the two image components and over the three seasons. The obtained time lag noticeably exceeds the value following for SBS 1520+530 from the classical model of optically thick geometrically thin AD. We considered two possible ways to resolve the discrepancy between the theory and observations. The first assumes an AD temperature profile flatter than the classical one. The second way is to consider an extended optically thick scattering envelope originated due to matter outflow from an AD interior.

\textbf{Conclusion:} Both explanations may be consequences of the super-Eddington accretion in SBS 1520+530. Using bolometric luminosity estimates available for SBS 1520+530 from the literature, we obtained a value of $\approx3.4$ for the Eddington ratio, which indeed indicates a moderately super-Eddington accretion regime.}


\footnotetext[1]{Accepted for publication in Astrophysics and Space Science}

\maketitle

\section{Introduction} \label{Introduc}   

Mass accretion onto a super massive black hole (BH) is presently believed to be the main mechanism to drive a huge luminous energy of active galactic nuclei (AGNs) and quasars in a broad region of spectrum, from the X-rays to optical and IR wavelengths. The maximum luminosity that can be generated by spherical accretion onto a BH with the mass $M_{BH}$ is

\begin{equation}
L_{Edd} = \dot M_{Edd} \cdot c^2 \approx 1.3\cdot10^{38} \cdot(M_{BH}/M_\odot) \mathrm{erg \cdot s}^{-1},      
\end{equation}
where $\dot M_{Edd}$ is the Eddington accretion rate, c the speed of light, and $M_{\odot}$ is the mass of the Sun. For the observed luminosity $L_{bol}$ produced by the mass accretion in real AGNs and quasars, a dimensionless quantity $\eta$ is introduced that represents mass-to-radiation conversion efficiency, so that $L_{bol} = \eta  L_{Edd} $. In transporting angular momentum  of the accreting matter outward, a disk is formed around the BH. The spatial structure of the disk and the emitted spectrum depend mainly on the rate of the matter inflow at the external boundary of the disk. Theories predict appreciable differences between the typical spectral energy distributions, geometries, temperature profiles and other emission properties of accretion disks (ADs), depending on a relationship between their mass accretion rates,  $\dot M$, and the corresponding Eddington accretion rate, $ \dot M_{Edd} \approx 1.7 \cdot 10^{17}(M_{BH}/M_\odot)$ g $\cdot$ s$^{-1}$. Slow-accreting systems, with accretion rates $\dot M < \dot M_{Edd}$, are expected to have thin accretion disks, for which the primary emission and observed properties are described by the standard thin AD theory by Shakura \& Sunyaev (1973). At high accretion rates, $\dot M > \dot M_{Edd}$, “thick” ADs are formed (Jaroszynski et al. 1980; Eggum et al. 1988; Szuszkiewicz et al. 1996; Fukue 2000). In 1988, Abramowicz et al. considered the intermediate case, $\dot M \approx \dot M_{Edd}$, and found out that the corresponding ADs, which they called “slim disks”, differ from both thin and thick disks in several astrophysically important aspects.

Due to the extreme remoteness of quasars and AGNs, the existing observational facilities do not allow obtaining their images with angular resolution that would be sufficient for examining their central regions. Therefore, various indirect methods are utilized to discern their spatial structure. Reverberation mapping (RM), initially proposed for determining distances between an ionizing central source and broad emission line (BAL) region, (Blandford \& McKee 1982; Peterson 1993; Wanders et al. 1997), has presently become a powerful tool for probing the temperature profiles of accretion disks and measuring the distances between their regions with different physical conditions, (e.g., Mudd et al. 2018, Yu et al. 2020; Sharp et al. 2024; Neustadt et al. 2024; Edelson et al. 2024; Marculewicz et al. 2024).

For gravitationally lensed quasars, another efficient tool for studying the spatial structure of their ADs has become available from observations after the discovery of microlensing phenomenon (Chang \& Refsdal 1979). Since the first detection of a microlensing event in Q2237+0305 (Irwin et al. 1989), a large number of such events have been detected in a set of gravitationally lensed quasars providing information about their structural peculiarities – in particular, AD sizes and their dependence on wavelength. Many microlensing studies (see e.g., Morgan et al. 2010; Blackburne et al. 2011; Jiménez-Vicente et al. 2012, 2014) have shown that the observed accretion disk sizes are on average a factor of 2–3 larger than those predicted by the standard thin disk model by Shakura \& Sunyaev (1973).

A similar discrepancy between the measured AD dimensions and those predicted by the standard thin disk model has also been demonstrated in many RM studies (e.g., Edelson et al. 2015; Fausnaugh et al. 2018; Pozo Nuñez et al. 2019). In our previous work (Berdina et al. 2021) we encountered the accretion disk size discrepancy mentioned above in analyzing the results of photometric RM applied to the Q2237+0305 light curves in filters $V$, $R$ and $I$: the measured inter-band time delays turned out to be noticeably larger than those predicted by the Shakura-Sunyaev thin AD model. To explain the discrepancy, we have admitted existence of a super-Eddington accretion regime in this object and shown that our results can be explained by the presence of a scattering envelope around the AD first considered by Shakura \& Sunyaev (1973).

In the present work, we used the archival monitoring data in filters $V$ and $R$ for a doubly lensed quasar SBS 1520+530 with the aim to determine the inter-band time delays and compare them with model predictions, as well as with those determined for the Q2237+0305 quasar in our previous work.

\section {Gravitationally lensed quasar SBS 1520+530 } \label{Gravlensed}   

For a quarter of a century since its discovery, the quasar SBS 1520+530 has not been the object of close attention from researchers. Below we provide brief information about this object, taken from the literature.
\begin{figure}[ht]
\centering
\resizebox{7.8 cm}{!}{\includegraphics{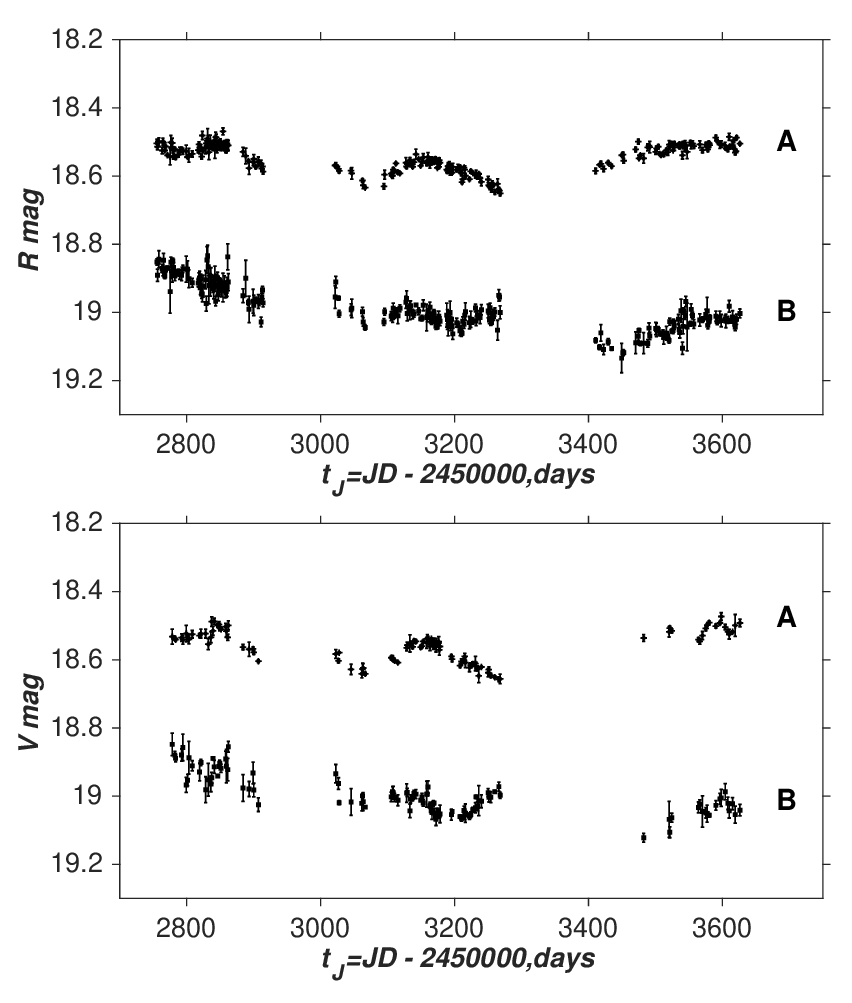}}
\caption{ Light curves of SBS 1520+530 in filters $R$ and $V$  for the 2003, 2004 and 2005 seasons from observations by Zheleznyak et al. 2003; Khamitov et al. 2006. The amplitude of the quasar intrinsic brightness variations in 2004 is 0.09 mag in filter $R$  and 0.1 mag in $V$  for the A component}
\label{light curves}
\end{figure}

This gravitationally lensed quasar is the BAL quasar at redshift $z_q$=1.855 from the Second Byurakan Survey resolved into two components by Chavushyan et al. (1997). The components with stellar magnitudes $V_A=18.^{m} 2$ and $V_B =18.^{m} 6$ (at the time of the object discovery), are separated by 1".56. The lensing galaxy with a redshift $z_{lens}$=0.765 is located 0."40 from the fainter of the two QSO images and is 0."12 offset from the line joining them (Crampton  et al. 1998). Auger et al. (2008) reported an updated value $z_{lens}$ = 0.761 obtained from the new spectroscopic data.

Monitoring campaigns (Burud et al. 2002;  Zheleznyak et al. 2003; Gaynullina et al. 2005;  Eulaers \& Magain 2011) have demonstrated noticeable intrinsic variability of the SBS 1520+530 quasar, which allowed to make the time delay estimates varying from 130 $\pm$ 3 days in Burud et al. (2002) to 125.8 $\pm$ 2.1 days in Eulaers \& Magain (2011). The presence of the quasar intrinsic brightness fluctuations has also made it possible to estimate inter-band time delay between the $V$ and $R$ light curves that turned out to be of 4.4 – 6.1 days in the observer’s coordinate plane (Koptelova \& Oknyanskij 2010).

We could find only two estimates of the BH mass for SBS 1520+530 in the literature, both made from emission line widths: $M_{BH} = 0.88 \cdot 10^9 M_\odot $ from the CIV line by Peng et al. (2006), and $M_{BH} = 0.4 \cdot 10^9 M_\odot $  by Ding et al. (2017) from the CIV, MgII, and H$\beta$ lines. Adopting the virial BH mass estimate by Peng et al. (2006), Morgan et al. (2010) used the scaling relationship log($r_{2500}/cm$)=(15.78$\pm$ 0.12) + (0.80$\pm$ 0.17)$\cdot$  log($M_{BH}/10^9 M_\odot$) to obtain $r_s \approx 5\cdot 10^{15}$cm  for $\lambda_{rest}$ = 250 nm. Microlensing variability, which is rather slow in this object, has made it possible to obtain only a few estimates of the accretion disk size: Khamitov et al. (2006) derived $r_s \approx  3.5 \cdot 10^{15}$ cm assuming the microlens velocity of 600 km $\cdot$ s$^{-1}$. Also, the SBS 1520+530 quasar is among 15 gravitationally lensed quasars for which Cornachione \& Morgan (2020) obtained the microlensing-based $r_{micr}$ and luminosity-based $r_{lum}$ estimates of the accretion disk size with the aim to constrain typical temperature profiles of accretion disks. Their estimates scaled to 2500{\AA} in the quasar rest frame are $r_{micr}=1.2\cdot 10^{16}$ cm = 4.6 light days and $r_{lum}=3.63\cdot 10^{15}$ cm = 13.2 light days – for measurements based on microlensing analysis and on thin disk predictions based on the observed luminosity, respectively.
\begin{table} [ht] 
\caption {General information about the data of observations of SBS 1520+530 used to determine the interband time delays in this work: duration of seasons, number of data points, and photometry errors (in $mag$) averaged over each season}
\label{data observations}
\centering
\begin{tabular}{lcccc}
\hline
  \multirow{2}{*}{Filter} & Season duration &  Number of data  & {Photometry error}  \\
 &(days) &  points &  (mag) &   \\
\hline
$R$, season 2003 &  159   & 79 & 0.0154  \\
$R$, season 2004 &  247   & 90 & 0.0104  \\
$R$, season 2005   & 216 & 70 & 0.0106  \\
\hline
$V$, season 2003 & 129  & 30 & 0.0168 \\
$V$, season 2004 & 246  & 61 & 0.0111 \\
$V$, season 2005   & 144 & 19 & 0.0150  \\
\hline
\end{tabular}
\end{table}

The black hole mass and AD size of SBS 1520+530 are very close to those of Q2237+030 – the object, for which we have earlier assumed a super-Eddington accretion regime (Berdina et al. 2021). Indeed, the values of $M_{BH}$ and $r_s$ for SBS 1520+530 indicated above lie well within the intervals of the existing estimates of these quantities for Q2237+0305 collected in tables 4 and 5 in the cited work by Berdina et al. (2021).  At the same time, Abolmasov \& Shakura (2012) obtained significantly different values of the accretion rate for these two objects, using an idea that presence of an optically thick scattering envelope resulting from matter outflow in the supercritical accretion regime is capable of breaking degeneracy between the BH mass and dimensionless accretion rate. This provided the dimensionless accretion rate for SBS 1520+530 almost twice as high as for Q2237+0305. Such a large difference for objects with almost the same black hole masses and accretion disk sizes seems rather strange and needs to be revisited with additional data and considerations.

\section{Initial data and processing} \label{Initdata}   

In the present work, we used the monitoring data for SBS 1520+530 taken with the 1.5-meter telescope of the Maidanak observatory in 2000-2005 (Zheleznyak  et al. 2003; Khamitov et al. 2006). Photometry is available from the site of the Institute of Astronomy of the Kharkov National University (http://www.astron.kharkov.ua/databases/index.html). The light curves in filters $V$ and $R$ for seasons of 2003-2005 are shown in Fig. \ref{light curves}, and Table \ref{data observations} contains a description of the observational data.

As is seen from Fig. \ref{light curves}, the 2004 light curves contain the largest number of data points and cover the longest time period. In addition, they include a fairly distinct fluctuation in the quasar’s brightness with the 0.1-magnitude amplitude, that makes them most promising for measuring the inter-band time delays. The $V$ and $R$ light curves of the 2003 and 2005 seasons consist of fewer data points and cover shorter time periods, but can also be considered suitable for measuring inter-band time lags, at least for component A. The light curves in filter $I$ are not shown in Figure 1 since they are all too sparse and therefore can hardly be used to determine the inter-band time delay for all seasons.

\begin{figure}[ht]
\centering
\resizebox{10.2 cm}{!}{\includegraphics{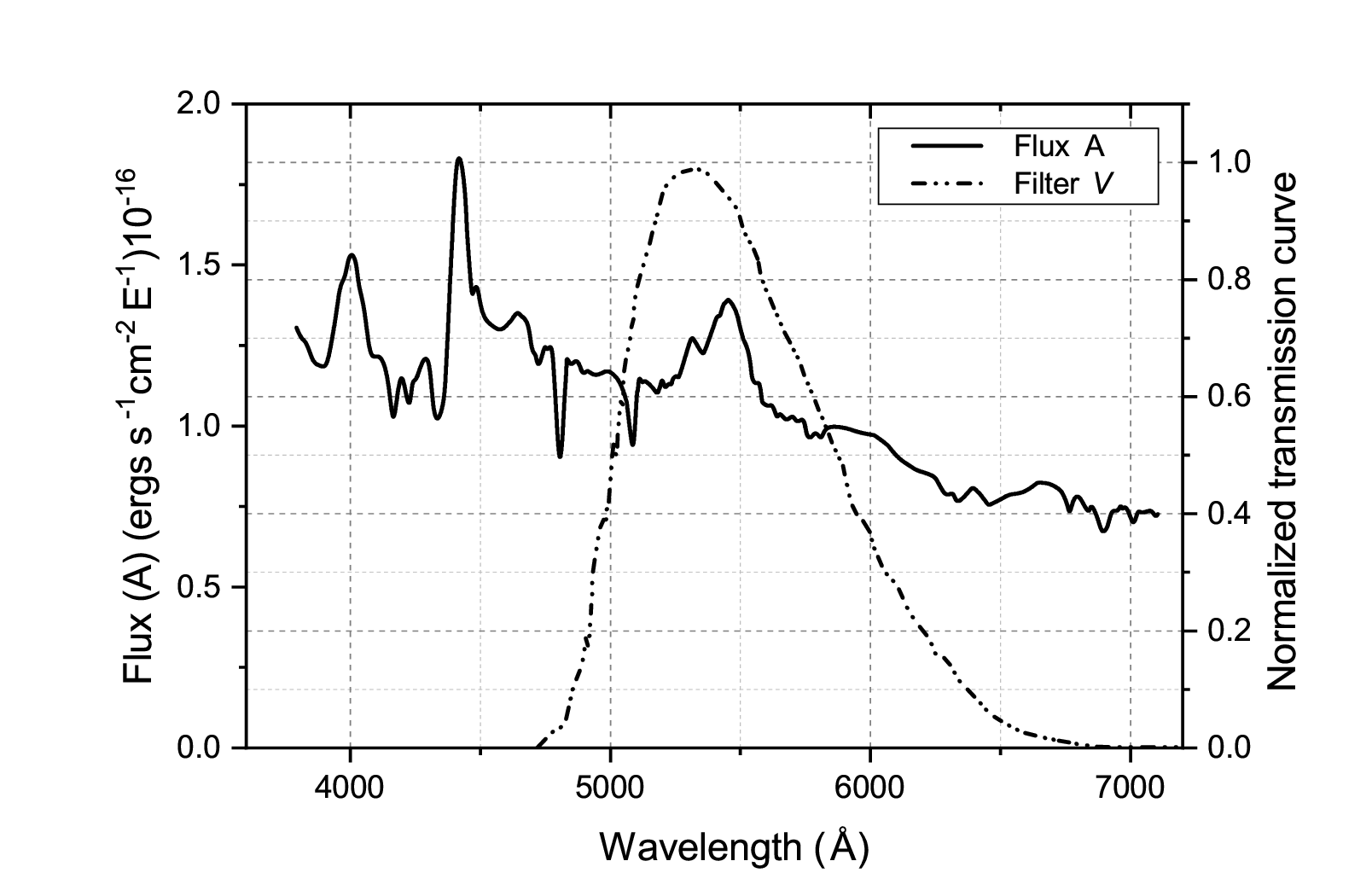}}
\caption{ The transmission curve of filter $V$ overlaid with the spectrum of the SBS 1520+530 (component A) taken from Chavushyan  et al. (1997)}
\label{contamination}
\end{figure}

The SBS 1520+530 spectrum obtained by Chavushyan et al. (1997), demonstrates several broad emission lines with one of them, AlIII/CIII], falling just in the middle of the transmission band of filter $V$. To visually illustrate the situation, we show in Fig. \ref{contamination} the transmission band of filter $V$ overlaid with the spectrum of SBS 1520+530. We estimated a corresponding contamination of the quasar light in continuum from this emission line and found it to be about 11\%, therefore, the light curve in filter $V$ can be considered as referring mostly to the continuum of the corresponding quasar region.

 In determining the inter-band time lags, we applied a method used in our previous works to measure the inter-band time lags in a quadruply lensed quasar Q2237+0305 and the gravitational time delays between the image components in other lenses (Tsvetkova et al. 2016,  Berdina \& Tsvetkova 2017, Berdina et al. 2021). In short, the method utilizes one of the useful properties of the orthogonal polynomials in representing the data of observations that allows, for example, eliminating or adding any term of a polynomial that approximates a particular light curve without necessity to recalculate the remaining expansion coefficients. A detailed description of the method can be found in (Tsvetkova et al. 2016).

\begin{figure} [ht] 
\centering
\resizebox{13.5 cm}{!}{\includegraphics{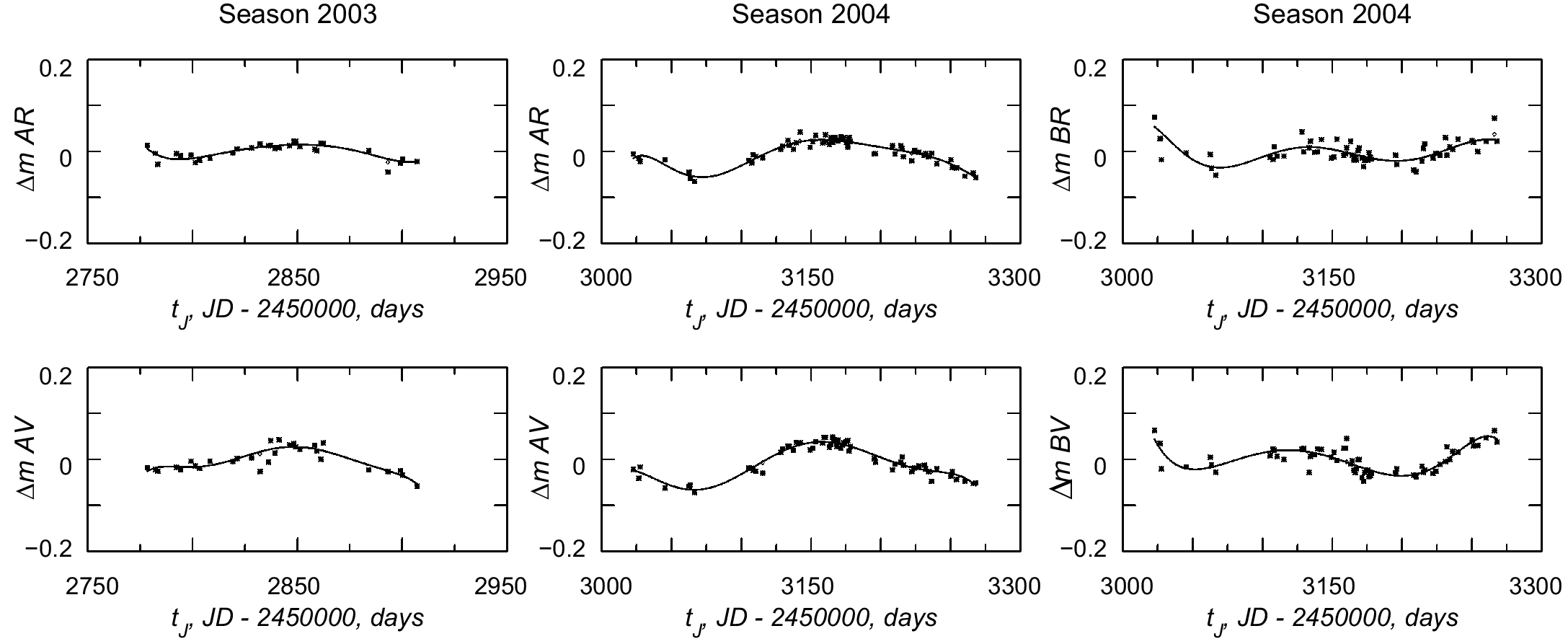}}
 \caption{ Polynomial approximations for the light curves in Fig. \ref{light curves}: the curves are reduced to the same magnitude level, and the first-order terms are eliminated ($ \Delta m $ is the brightness variations in magnitude relative to the average level for a given season)}
\label{polynomial approximations}
\end{figure}

In Fig. \ref{polynomial approximations}, approximations of the initial light curves (Fig. \ref{light curves}) by series expansion in orthonormalized Legendre polynomials are shown (solid curves). The light curves are reduced to the same magnitude level, and the first-order terms are eliminated. The further processing consists in calculation of the cross-correlation functions for pairs of light curves represented by their polynomial  approximations  in the evenly sampled data points. To exclude edge effects emerging in calculation of cross-correlation function, we used a cross-correlation procedure similar to the Locally Normalized Discrete Correlation Function, (LNDCF), proposed by Lehar et al. (1992) . The value of the current time shift between the compared functions, at which the cross-correlation function reaches its maximum, is taken as an estimate of the desired inter-band time delay.

The estimates of the rest-frame values of the inter-band time delay ($\Delta t$=$\Delta \tilde t/(1+z_q)$) between the SBS 1520+530 quasar brightness fluctuations in filters $V$ and $R$ are presented in Table \ref{rest-frame time}, where the corresponding confidence intervals $CI$ for the 90\% confidence level are also indicated. The corresponding rest-frame effective wavelengths for these filters are $\lambda_{R}$ = 222.4nm, and $\lambda_{V}$ = 191.1nm.

Positive time delays mean that the light curve corresponding to the first letter in a $\Delta t$ subscript lags that one corresponding to the second symbol, for example, an entry  $\Delta t_{RV}=1.22\pm0.71 $ in Table \ref{rest-frame time} means that the $R$-band light curve lags behind the $V$-band light curve by 1.22 days. For season 2004, the light curves of both A and B images have been used to provide the results presented in column 3. For seasons 2003 and 2005, however, there was no possibility to obtain reliable estimates of the inter-band time lags for component B.

To estimate the uncertainties of measuring the time lags, we compared the results of calculation obtained from realizations of the initial light curves subjected to the following modifications. Several randomly selected data points (from a few up to  30\% of the total amount in each light curve) are successively excluded from the
original light curves thus forming new separate subsamples of the same process. The procedure described above is applied further to each subsample to provide new polynomial approximations of a light curve, and new lag estimates are obtained. A number of such trials reached 20 for each pair of light curves, and then a set of such estimates was used to obtain the average lag values for every component and season. In doing so, we proceeded from the fact that the estimates of statistical parameters of a stationary random process are equally valid both from the analysis of the whole signal record, and by averaging the results of processing separate realizations (subsamples) of the process. The RMS deviation of individual estimates from the average was taken as an error of our  time delay estimates that turned out to be $\pm 0.71$, $\pm 0.68 $, and $\pm 0.59$ for image A in seasons 2003, 2004 and 2005, respectively, and  $\pm 0.41$ for image B. It should be noted that the above procedure chosen to estimate uncertainties of determining the lags resembles in some features the FR/RSS method proposed by Peterson et al. (1998).

\begin{table}  [ht]
\caption {Rest-frame time delays $ \Delta t $ in days of the SBS 1520+530 quasar for the $RV$ pairs of light curves and the corresponding confidence intervals $CI$ for the 90\% confidence level. Positive time delays correspond to the $R$ light curves lagging the light curves in the $V$ band  }
\label{rest-frame time}
\centering
\begin{tabular}{lccc}
\hline
 Season&  2003&  2004&  2005 \\
\hline
$ \Delta t_{RV}(A)$ & $1.22\pm0.71$  &$1.66\pm0.68$ & $1.21\pm0.59$ \\
$CI$      & (0.66; 2.51) & (1.29; 3.08) & (0.83; 2.39) \\
\hline
$\Delta t_{RV}(B)$  & -- &$0.91\pm0.41$ & -- \\
$CI$& -- & (0.65; 1.71)  & --\\
\hline
$\Delta t_{RV}(A+B)$  & -- &      $1.25 \pm 0.63$   &      --\\
$CI$& -- & (0.84; 2.48)  & --\\
\hline
\end{tabular}
\end{table}

\begin{figure} [ht]
 \centering
\resizebox{12.5 cm}{!}{\includegraphics{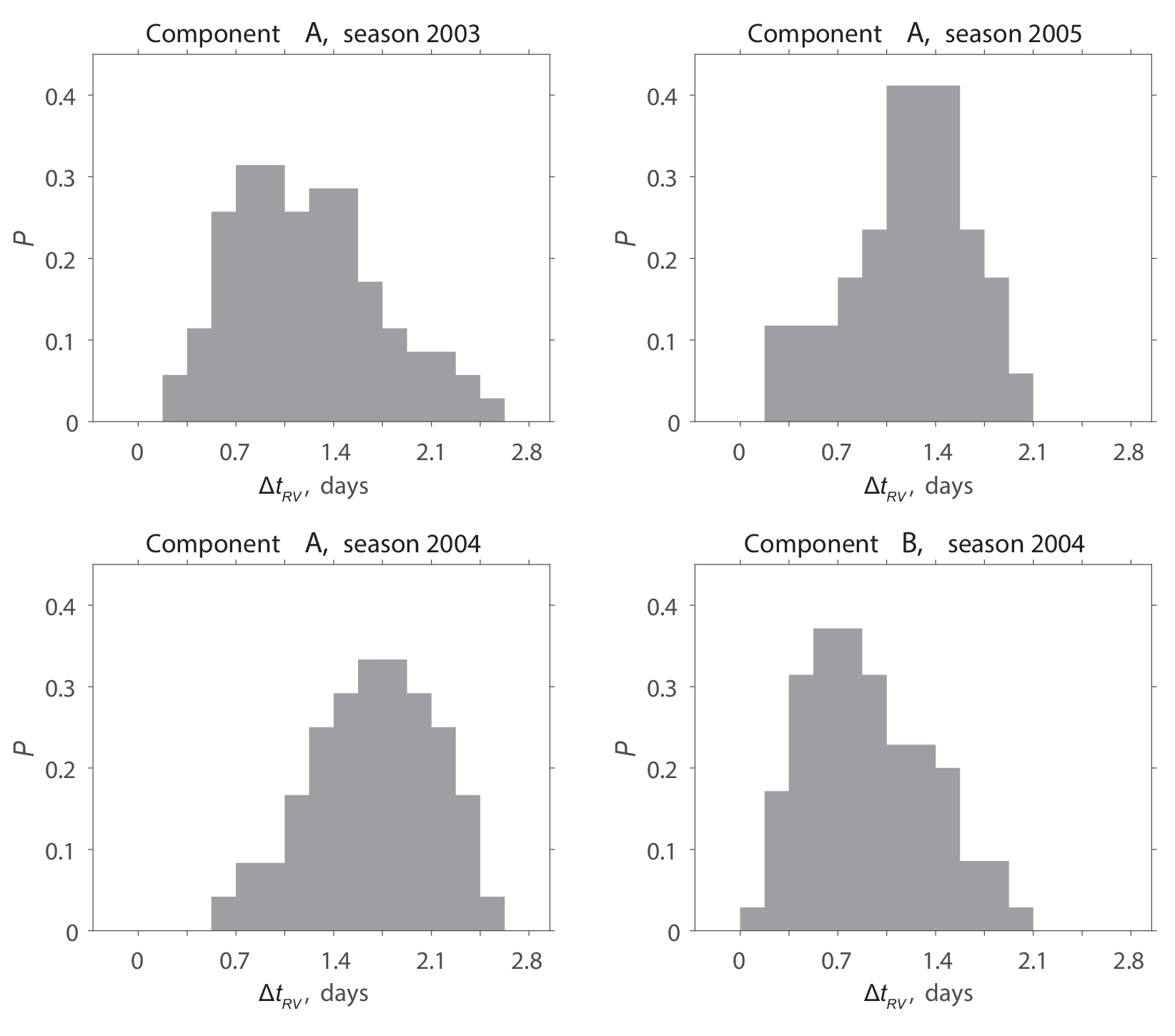}}
 \caption{ Histograms of the probability density distribution $P$ for the estimates of the interband time delays $\Delta t_{RV}$, presented in Table \ref{rest-frame time} }
 \label{histograms}
\end{figure}

In Fig. \ref{histograms}, we demonstrate the histograms of the probability density distribution $P$ for the estimates of inter-band time delays $\Delta t_{RV}$ for seasons 2003-2005 presented in Table \ref{rest-frame time}. Each histogram is built from the results of processing, for a specific image component and season, the entire totality of the corresponding pairs of light curves modified as described above. The probabilities $P$ are taken to be proportional to the frequencies of falling the individual estimates within the corresponding 0.175-days intervals of the time lags. The histograms have fairly clear maxima and visualize the reliability of our determinations of the inter-band time lags uncertainties. They also confirm the closeness of the $\Delta t_{RV}$ estimates for the 2003 and 2005 seasons based on the light curves of component A, and at the same time seem to favor the reality of the difference between the results for components A and B for the 2004 season. Below, we will consider the value $< \! \Delta t_{RV} \! >$ = 1.25 $\pm$ 0.63 days, that is the average over all seasons and components.

As is seen from Table \ref{rest-frame time}, differences between the time lags of components A and B for the same season, as well as of component A in the different seasons are large enough, so that a question might arise if  averaging over all seasons and components makes sense. In general, the answer may be negative. Indeed, a class of mechanisms of quasar variability grouped as “accretion disk instabilities” – varying mass accretion rate, flares or hot spots, avalanches  (Kawaguchi 1998; Wilhite et al. 2004; Ross et al. 2018), – admits certain structural changes in accretion disks. If this were the case, and if physical timescales of these processes were comparable to those of reverberation responses, the light curves of the same component in different seasons should be ascribed to different episodes of quasar variability and, therefore, different estimates of the reverberation time delays might be expected. Similarly, in systems with significant gravitational time delays between components, such as SBS 1520, it is conceivable that reverberation time lags measured for different components - even within the same observing season - may differ. Thus, the scatter of estimates in Table \ref{rest-frame time} can be caused by both the effects of the factors mentioned above and the errors inherent in the original light curves and in methods of their processing aimed at obtaining the lags. The available data are insufficient to make a confident conclusion, but nevertheless, in order not to mask the possible influence of these factors on the time delay estimates, we processed the light curves for each season separately, without resorting to the formation of a longer common realization.

\section {Comparison with the classical thin-disk model}  \label{Comparison}  

In the classical model of the optically thick geometrically thin AD by Shakura \& Sunyaev (1973), the observed emission of the whole disk is assumed to be a superposition of the blackbody radiations generated in the disk locally at each particular radius. The flux per unit disk area of the energy that viscously dissipates in matter accretion at radius $r$,

\begin{equation}
D(r)=\frac{3GM\dot M}{8\pi r^3 },
\end{equation}
is converted into a blackbody radiation so that

\begin{equation}
\sigma T^4=\frac{3GM\dot M}{8\pi r^3 }.
\end{equation}

This relation allows to express the disk scale radius at an effective wavelength $\lambda_0 $, where the disk temperature satisfies a condition $kT=hc/\lambda_0$:

\begin{equation}
\label{rlamda0}
r(\lambda_0)=\left[\frac{3GM\dot M \lambda_{0}^4 k^4}{8\pi \sigma c^4 h^4}\right]^{1/3}.
\end{equation}
Here, $G$ is the gravitational constant, $\sigma $, $k$ and $h$ are the Stefan-Boltzmann, Boltzmann and Planck constants, respectively, and $c$ is the speed of light.

Thus, in the thin disk model the radius scales with wavelength as $r(\lambda) \sim  \lambda^{4/3}$. Such scaling means that the shorter-wavelength radiation emerges closer to the disk center. For variable sources, variations of the high-temperature radiation from the disk interior are often observed to precede those in longer wavelengths emitted at larger radii. Such time lags detected in many RM studies provide valuable information about the AD structure unattainable from observations which imply imaging. To explain these observed time lags, Krolik et al. (1991) suggested that the hard radiation from the inner region of the disk, when propagating outward, is reprocessed into optical photons in the outer and therefore colder AD regions. Qualitatively, our results shown in Table  \ref{rest-frame time} are consistent with the thermal reprocessing scenario by Krolik et al. (1991).

After substituting the numerical values of physical constants $\sigma $, $k$, $h$  and $G$ and introducing the dimensionless quantities $(\lambda/\mu m)$,  $M_{BH}/10^9 M_\odot$, and $(L/ \eta L_{Edd})$, equation (\ref{rlamda0}) will have the form more convenient for calculations and often used by other authors (e.g. Poindexter \& Kochanek 2010; Morgan et al. 2010; Mudd et al. 2018):

\begin{equation}
\label{equation 9.7}
r_\lambda=9.7 \cdot 10^{15}\left(\frac{\lambda}{\mu m}\right)^{4/3} \left(\frac
{M_{BH}}{10^9 M_\odot}\right)^{2/3}\left(\frac {L}{\eta L_{Edd}}\right)^{1/3}.
\end{equation}
Here, $L$ and $L_{Edd}$ are the disk luminosity and Eddington luminosity ($L_{Edd}=10^{38}\cdot(M_{BH}/M_\odot)$ erg$\cdot$s$^{-1}  $ ),  and $\eta$ is the accretion efficiency. When there is no other information, $\eta$ is accepted to be 0.1, and the disk luminosity $L$ is often assumed to equal $L_{Edd}$, thus providing the upper limit to the estimate of $r_\lambda$.

From Eq. (\ref{equation 9.7}), one can calculate the values of the disk radii for the rest-frame wavelengths corresponding to the effective wavelengths of the filters used. Adopting for the SBS 1520+530 black hole mass the estimate by Peng et al. (2006), $M_{BH}=8.8\cdot 10^8 M_\odot$, we obtain the following values of the AD radii predicted by the thin disk model for filters $V$ ($\lambda_{rest}(V)$ = 191.1 nm) and $R$ ($\lambda_{rest}(R)$ = 222.4 nm):

\begin{equation}
\label{equation 3.04}
r_{\lambda}(V) = 2.11 \cdot 10^{15} (cm); \;\: \:\:  r_{\lambda}(R) = 2.59 \cdot 10^{15} (cm),
\end{equation}
or, reducing the scale length (\ref{equation 9.7}) to the half-light radius, $r_{1/2} =2.44  \cdot r_{\lambda} $:

\begin{equation}
\label{half}
r_{1/2}(V) = 5.14 \cdot 10^{15} (cm); \;\: \:\:  r_{1/2}(R) = 6.32 \cdot 10^{15} (cm).
\end{equation}

The SBS 1520+530 quasar enters a sample of 11 gravitationally lensed quasars for which Morgan et al. (2010) published the estimates of AD radii obtained in microlensing analysis of multi-band monitoring data. The authors give $r_{\lambda} = 5.01 \cdot 10^{15}$ cm at $\lambda_{rest}$ = 245 nm for the SBS 1520+530 AD in their table 1. Reducing this radius to $\lambda_{rest}(V)$  and $\lambda_{rest}(R)$  according to the thin-disk scaling $r(\lambda_2) = r(\lambda_1) (\lambda_2 / \lambda_1)^{4/3}$, we obtain for the disk half-light radii at the effective wavelengths of filters $V$ and $R$ the following values based on observations:

\begin{equation}
\label{filters}
r_{1/2}(V) = 8.78 \cdot 10^{15} (cm); \;\: \:\:  r_{1/2}(R) = 10.76 \cdot 10^{15} (cm).
\end{equation}

We see an illustration of the AD size discrepancy mentioned in the Introduction: the radii based on microlensing observations (\ref{filters}) are larger than the theoretic ones (\ref{half}) predicted by the classical thin disk model for the black hole mass $M_{BH}=8.8\cdot 10^8 M_\odot$.

Even larger discrepancy takes place between the microlensing-based AD sizes and their estimates calculated from the observed luminosities in the framework of the thin disk model assuming blackbody emission within each annulus of the disk. This has been demonstrated by Cornachione \& Morgan (2020) for a sample of 15 gravitationally lensed quasars (with the SBS 1520+530 quasar among them). Their table 1 contains both microlensing-based and luminosity-based size estimates, with the latter being systematically lower for all objects but one.

As is noted in the Introduction, a similar discrepancy between the theoretically expected AD sizes and those determined from observations follows from some reverberation mapping (RM) studies. RM is known to provide light travel times between the accretion disk regions with different physical conditions. To interpret the observed time lags between the light curves at different spectral bands, a simplified scheme for emerging reverberation responses is used sometimes (e.g., Mudd et al. 2018; Berdina et al. 2021). In short, for a centrally illuminated accretion disk, reverberation responses are supposed to be formed at each spectral band in some annular zones. The zones are located at the distances from the central source where the blackbody temperatures match the passbands of the corresponding filters. In their propagation towards the disk periphery, initial fluctuations of the hard radiation from a central region are reprocessed into the longer-wavelength signals, with the time lags determined by the proper light travel times between the zones, and with distortions, which are due to sizes, shapes, and positions of the re-emitting regions. In the framework of this simplified scenario, supposing the disk viewed face-on, the time lag $\Delta t_{RV}$ between the light curves in a particular filter pair ($V$ and $R$ in our case) is:

\begin{equation}
\label{equation tau}
\Delta t_{RV} = (r_R - r_V)/ c,
\end{equation}
where $c$ is the speed of light.

Eq.\ref{equation tau} provides a very simple relation between the uncertainty of our measurements of $\Delta t_{RV}$ and the resulting uncertainty in a distance $(r_R - r_V)$  between the disk zones responsible for the observed lag. The RMS error in $\Delta t_{RV}$ of $\pm 0.63$ days, as indicated in Table \ref{rest-frame time}, is equivalent to the uncertainty in $(r_R - r_V)$ constituting  $\pm 1.5 \cdot 10^{15} $ cm, or approximately  $\pm 0.8 \cdot 10^{15} $ cm in radius.

Applying (\ref{equation tau}) to the radii $r_{1/2}(V)$ and $r_{1/2}(R)$ predicted by the classical thin disk model and indicated in (\ref{half}), we obtain $\Delta t_{RV}$ = 0.44 days for the expected time lag between the $R$ and $V$ light curves. It is much less than the time lags determined in this work (see Table \ref{rest-frame time} and the last paragraph of Sec. 3). Similarly, applying (\ref{equation tau}) to the half-light radii (\ref{filters}) reduced from the microlensing data at $\lambda_{rest}$ = 245 nm (Morgan et al. 2010) to the $V$ and $R$ effective wavelengths according to the thin-disk scaling $r(\lambda) \sim  \lambda^{4/3}$, we obtain $\Delta t_{RV}$ = 0.76 days. It is closer to our estimates but still less than even the minimal value $\Delta t_{RV}$ = 0.91 days indicated in Table \ref{rest-frame time} for component B in 2004.

\begin{table} [!htp]
\caption {Values of exponents in the AD size-wavelength scaling $r \sim  \lambda ^{1/\beta}$ derived from microlensing data (from literature for 2008-2018), and the corresponding exponent values in temperature profiles $T \sim  r^{ - \beta} $}
\label{betta}
\centering
\begin{tabular}{lcccl}
\hline
 Object &  $M_{BH}/10^9 M_\odot$ &  $p=1/\beta$ in   & $\beta$  in $T \sim   r^{ - \beta} $& References  \\
         &               &           $r\sim \lambda ^p $ &       &          \\
\hline
&       &  $0.77 <\beta <2.67$ & $0.37 <1/\beta <1.3 $ &  Bate et al. 2008$^{*}$ \\    
MG0414+0534& 1.82\footnotemark[1]   & $1.8\pm 0.6$ & 0.55  & Bate  et al. 2018\\
 & 2.51\footnotemark[2]  &  $1.5\pm 0.84$  &  0.67  & Blackburne et al. 2011\\
 \hline
 &   &  $0.67 \pm 0.55$ & 1.49 &  Blackburne  et al. 2011 \\
HE0435-1223 & 0.5\footnotemark[1]  & $0.75-1.3 (\pm0.6)$ & $1.33-0.77 $& Jimenez-Vicente et al. 2014\\
 &  & $1.3\pm 0.3$& $0.77$ & Mosquera et al. 2011\\
\hline
WFIJ2026-4536&  1.0\footnotemark[3]   & $2.3\pm0.5$ & $0.43 $& Bate et al. 2018 \\
 &  & $0.27\pm 0.53$& $3.7$ & Blackburne et al. 2011 \\
\hline
 &  & 1.64-1.12 & 0.61-0.89 & Poindexter et al. 2008 \\
  & 1.18\footnotemark[2]  & $1.0\pm 0.5$& 1.0 & Blackburne et al. 2015 \\
HE1104-1805 &  0.59\footnotemark[4]  & $0.75-0.9 (\pm 0.4)$  & 1.33-1.11& Jimenez-Vicente et al. 2014 \\
   &  2.37\footnotemark[5]   & $0.7\pm 0.1$& $1.43$ & Motta  et al. 2012 \\
  &  & $1.1\pm 0.6$& $0.91$ & Muñoz  et al. 2011 \\
\hline
& 1.95\footnotemark[5] & 0.9-1.0 & 1.11-1.0& Mediavilla et al. 2011 \\
 SBS0909+532  & 3.87\footnotemark[1]  & $0.75-0.9 (\pm 0.4)$  & $1.33-1.11$ & Jimenez-Vicente  et al. 2014 \\
 & 1.29\footnotemark[2]  &   &  &  \\
\hline
SDSSJ1004+4112 &  2.02\footnotemark[5] &  $1.1\pm 0.4$ & 0.9 & Motta  et al. 2012\\
 & 0.39\footnotemark[6]   & $1.0-1.3 (\pm 0.6)$  & $1.0-0.77$ & Jimenez-Vicente  et al. 2014 \\
\hline
SDSSJ0924+0219 & 0.11\footnotemark[1]  &  $<1.34 $ & >0.75 & Floyd et al. 2009$^{*}$ \\
 &   & $0.17\pm 0.49$  & $5.88$ & Blackburne et al. 2011 \\
\hline
 & 0.9\footnotemark[6]  &  $1.2\pm 0.3$ & 0.83 & Eigenbrod  et al. 2008 \\
 Q2237+0305 &  0.47\footnotemark[5]  & $0.5-0.6 (\pm 0.4)$  & $2.0-1.67$ & Jimenez-Vicente  et al. 2014 \\
 &  1.2\footnotemark[4]  &   &   &   \\
 \hline
B1422+231 & 4.79/2.23\footnotemark[1] &  $1.4\pm 0.5$ & 0.71 & Bate  et al. 2018 \\
& 7.94\footnotemark[2]  &  &  &  \\
\hline
HE0230-2130 &    &  $-0.56\pm 0.47$ & ???  & Blackburne  et al. 2011 \\
\hline
RXJ0911+0551 & 0.8 \footnotemark[1] &  $0.17\pm 0.41$ & 5.88 & Blackburne  et al. 2011 \\
\hline
HE1113-0641 &   &  $0.05\pm 0.49$ & 20 & Blackburne et al. 2011 \\
\hline
 &  0.92/1.23\footnotemark[1]  &  $0.40\pm 0.45$ & 2.5 & Blackburne  et al. 2011 \\
 PG1115+080  &  0.63\footnotemark[2]  &    &    &   \\
    &  1.23\footnotemark[6]  &    &    &   \\
\hline
RXJ 1131-1231 & 0.06\footnotemark[1] &  $0.40\pm 0.50$ & 2.5 & Blackburne et al. 2011 \\
\hline
WFIJ2033-4723&    &  $-0.63\pm 0.52$ &  ???   & Blackburne et al. 2011 \\
\hline
HE0047-1756 & 1.48\footnotemark[1]  &  $2.3\pm 0.8 $ &0.43 & Rojas  et al. 2014 \\
\hline
SDSSJ1155+6346 &  &  $1.5\pm 0.6 $ &0.67 & Rojas et al. 2014 \\
\hline
HE0512-3329 &  &  $1.25-1.4 (\pm 0.6) $ & 0.8-0.71 & Jimenez-Vicente et al. 2014 \\
\hline
 &  2.01\footnotemark[5] &  $0.25-0.9(\pm 0.7) $ & 4.0-1.11 & Jimenez-Vicente  al. 2014 \\
 QSO0957+561  & 0.72\footnotemark[4]   &   &   &  \\
  & 1.42\footnotemark[2]   &   &   &  \\
\hline
SDSSJ1029+2623 &  &  $0.0-0.9 (\pm 0.7) $ & $>1.11$ & Jimenez-Vicente et al. 2014 \\
\hline
\end{tabular}
\footnotetext{Notes: Exponents in $r \sim  \lambda ^{1/\beta} $ were derived from microlensing data; asterisk over two references mean single-epoch imaging.}
\footnotetext{References: (1) Peng  2006; (2) Greene et al. 2010; (3) Bate et al. 2018; (4) Assef et al. 2011; (5) Mosquera \& Kochanek 2011; (6) Morgan et al. (2010).   }
\end{table}

Thus, the AD radii (\ref{half}) in filters $V$ and $R$ calculated in the framework of the classical model of a geometrically thin disk by Shakura and Sunyaev (1973) do not provide the values of the reverberation time delay $\Delta t_{RV}$ consistent with our measurements.

\section{SEDs, temperature profiles and the AD size discrepancy}  \label{Effsuper-Eddington}   

Discrepancies between the AD sizes predicted by the classical thin disk model of Shakura and Sunyaev and those obtained from observations have led to emergence of a number of alternative disk models. Various AD models use a generalized form $T(r) \sim  r^{ - \beta} $  for the effective surface temperature profile, which is equivalent to the AD radius varying in wavelengths according to the power-law relationship $r_{\lambda} \sim \lambda^{p} $ , (spectral profile hereafter), with $p=1/\beta$. The value of the temperature profile slope $\beta $ or, equivalently, the slope of a spectral profile $p=1/\beta$, is controlled by mechanisms of generation and transportation of the accretion energy and in turn, strongly affects the energy distribution in the overall disk spectrum. The temperature profile slopes vary in different models from $\beta \approx 0.5$  for the slim-disk model by Abramowicz et al. (1988) to, e.g., $\beta \approx 7/8$ (Agol \& Krolik 2000), with $\beta \approx 3/4$  for the Shakura \& Sunyaev (1973) model of an optically thick geometrically thin accretion disk.

While reverberation mapping of AGNs and quasars, as well as multiband observations of microlensing events provide the AD sizes mostly larger than those predicted by the classical thin disk model, measurements of the AD spectral structure parameters $\beta$ and $p=1/\beta$ give fairly divergent results. Table \ref{betta} demonstrates the diversity of estimates of parameters $\beta$ and $p=1/\beta$ made for 20 gravitationally lensed quasars from microlensing observations we could find in the literature. Some of them favor the temperature profiles shallower than the thin AD prediction, $\beta<3/4$, while most show the slopes close to the thin-disk value $\beta=3/4$ or steeper.

\begin{figure} [!htp]
\resizebox{13.6 cm}{!}{\includegraphics{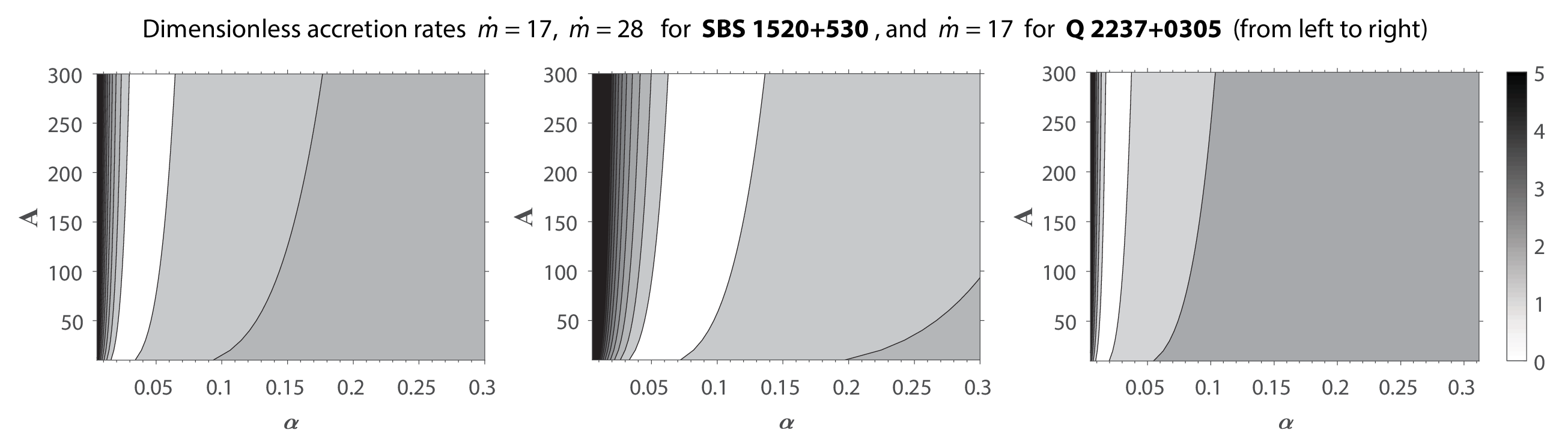}}
 \caption{Maps of differences between the interband time lag $< \! \Delta t_{RV} \! >$ = 1.25 days determined in the present work and calculated according to equations (\ref{equation teff})  and (\ref{equation ropt}) . The absolute values of differences between these quantities are presented in grayscale (shown on the right side of the figure), with the lightest regions corresponding to the smallest differences. The maps are built for two dimensionless accretion rates for SBS 1520+530 and one for Q 2237+0305, from left to right: $\dot m = 17$, $\dot m = 28$, and $\dot m = 17$}
\label{maps all masses}
\end{figure}

The shape of the spectral energy distribution (SED) of the overall radiation from quasars and AGNs is another important physical characteristic available from observations that is closely related to the AD spectral structure parameters. The disk spectrum can be obtained by integration over the radius of black-body radiation produced at each particular radius locally, thus providing the shape of the resulting SED that depends only on the AD temperature profile (Shakura \& Sunyaev 1973):

\begin{equation}
\label{finu}
F_{\nu}  \sim \nu ^{ \gamma} \sim \nu ^{(3-2/ \beta)}.
\end{equation}

It is seen that with such a dependence of the SED shape on the temperature profile slope $\beta$, the scatter of the $\beta$ values similar to that shown in Table \ref{betta}  may result in unacceptable scatter of the spectral index $\gamma$. Given such large uncertainties in determining the AD structure parameters from observations and the lack of an AD model capable of resolving the discrepancies between theory and observations, Gaskell (2008) proposed to inverse the problem and determine the AD temperature profiles from the observed spectra, in particular, from their spectral indices in the UV/optical range. With the UV/optical spectral index of a composite spectrum $\gamma$ = – 0.54 obtained by Gaskell et al. (2004) for a sample of 72 radio-loud and 1018 radio-quiet AGNs, relation (\ref{finu}) implies a temperature profile slope of $\beta$ = 0.57.

To date, a large number of independent estimates of the UV/optical spectral indices have been accumulated for the composite SEDs of various samples of AGNs and quasars, which vary between $\gamma$ = – 0.32 (Francis et al. 1991) and $\gamma$ = – 0.99 (Zheng et al. 1997) through a number of intermediate values, providing, according to (\ref{finu}), $0.52 < \beta < 0.6$ for the temperature profile slope. As is seen, in spite of a rather wide spread in the estimates of $\gamma$ explained mainly by difficulties in accounting for reddening, the corresponding values of $\beta$ are steadily less than the classical thin-disk $\beta$ = 0.75, and consistent with the temperature profile slope $\beta$ = 0.57 found by Gaskell (2008).

The limiting case opposite to the mechanism of local energy generation, is the disk irradiation by a single central source of thermal radiation. Heating at longer radii will then result from the outward heat flow from the hotter interior regions. Since the flux from a central source falls as $r^{-2}$, the effective black-body temperature will decrease as
\begin{equation}
\label{tandr}
T \sim (r^{-2})^{1/4} \sim r^{-1/2}.
\end{equation}

With the power-law index in a temperature profile of $\beta$ = 0.5 (and thus, with the AD radius varying in wavelength as $r_{\lambda}\sim \lambda^2$), the shape of the overall AD spectrum will be, according to (\ref{finu}), $F_{\nu}  \sim \nu ^{ -1}$. This is the so-called “slim” disk of Abramowicz et al. (1988) we have already mentioned above. It is interesting to note that $\beta$ = 3/4 following from the classical thin-disk theory, leads to a SED of the form $F_{\nu}  \sim \nu ^{+1/3}$ that has never been observed in the UV/optical spectral range (Gaskell 2008).

Of course, purely local production of the black-body radiation in matter accretion, as well as illumination by a central source of thermal radiation alone are both idealizations. In reality, some other processes can also be involved in the accretion process in disks, which can affect the AD structure parameters and thus can be potentially useful for solving the problem of the discrepant AD sizes. These may be, e.g., scattering a significant fraction of the AD emission at larger radii (Morgan et al. 2010), flattening a disc temperature profile because of some reasons (Dai et al. 2010; Morgan et al. 2010; Bonning et al. 2013), effects of an inhomogeneous AD with large temperature fluctuations (Dexter \& Agol 2011; Cai et al. 2018), matter outflow from the disk interior (disk winds), ( Fukue 2004; Slone \& Netzer 2012), a super-Eddington accretion disc with wind producing an extended optically thick envelope (Nishiyama et al. 2007; Abolmasov \& Shakura 2012; Li et al. 2019; Kitaki et al. 2021), and others.

The assumption that the temperature profile shallower than the classical $T(r) \sim r^{-3/4}$ is capable to reduce discrepancy between the observed and theoretically predicted AD sizes has been repeatedly reported by many authors, (e.g., Morgan et al. 2008, 2010; Hall et al. 2014; Poindexter et al. 2008; Bate et al. 2018; Dai et al. 2010). Recently, Cornachione \& Morgan (2020) reported that a temperature profile slope of $0.37 < \beta < 0.56 $ is found to be capable of resolving discrepancy between the microlensing-based and luminosity-based AD sizes. They noted consistency of their result with the work by Gaskell (2008), where the slope $\beta$ = 0.57 has been derived from the shape of a composite quasar spectrum. They also noted that their result confirms inferences of a phenomenological model of an AD with wind proposed by Li et al. (2019).

For a thin AD with wind, their model considers the inward power-law decrease of the accretion rate, $\dot M(r)\sim r^s$, where $s$ is the wind strength, and incorporates the angular momentum transfer by the wind. For a thermally radiating black-body disc with the accretion rate radial profile of $\dot M(r)\sim r^s$, the effective temperature profile $T(r) \sim r^{-\beta}$ will then have a slope of $\beta = (3-s)/4 $. A very similar model of an AD with wind is proposed by Sun et al. (2019) in their work published almost simultaneously with Li et al. (2019). They also admit significant matter outflow with a power-law dependence of the mass accretion rate $\dot M(r)\sim r^s$ resulting in the same temperature profile slope of $\beta = (3-s)/4 $.

Table 1 in the work by Li et al. (2019) contains the values of $s$, which provide consistency between the microlensing-based and luminosity-based AD sizes for 9 gravitationally lensed quasars from the list of 11 objects by Morgan et al. (2010). For SBS 1520+530, the value of $s$ = 1.1 is indicated resulting in the temperature profile slope $\beta$ = 0.475 that falls within the interval $0.37 < \beta < 0.56$ obtained by Cornachione \& Morgan (2020) and does not contradict the slope value of $\beta$ = 0.57 determined from the UV/optical spectral index in a composite quasar spectrum, as proposed by Gaskell (2008).

By repeating simple calculations similar to those described in the previous section to determine the time lags expected for $r_{\lambda}= 5.01 \cdot 10^{15}$  cm at $\lambda_{rest} $ = 245 nm from Morgan et al. (2010) with the thin-disk scaling $r(\lambda) \sim \lambda^{4/3}$, we must now reduce this radius to $\lambda_{rest}(V)$  and $\lambda_{rest}(R)$ according to the Li et al. (2019) scaling $r(\lambda_2) = r(\lambda_1) (\lambda_2 /\lambda_1)^{4/(3-s)}$ to obtain the disk radii at the effective wavelengths of filters $V$ and $R$. Adopting $s$ =1.1 indicated for SBS 1520+530 by Li et al. in their table 1, we will have for the $V$ and $R$ half-light radii:
\begin{equation}
\label{radii}
r_{1/2}(V)= 7.25 \cdot 10^{15} (cm); \;\: \:\:   r_{1/2}(R)= 9.98 \cdot 10^{15} (cm).
\end{equation}

This provides, according to (\ref{equation tau}), $\Delta t_{RV}$=1.05 days that is less than our estimate $\Delta t_{RV}=1.25\pm0.63$ days, though lies within the indicated error bar. A more accurate match to our $\Delta t_{RV}$=1.25 days can be obtained with the wind strength of $s\approx1.5 $ that corresponds to the temperature profile slope of $\beta$=0.375.

Thus, our estimate of the inter-band time lag for SBS 1520 +530 can be reproduced by a shallower-than-thin-disk temperature profile with a power-law index that is close to the lower boundary of the range $0.37<\beta<0.56$, proposed by Cornachione \& Morgan (2020) to reconcile the microlensing-based and luminosity-based AD sizes. Interestingly, about 25\% of multicolor microlensing observations presented in Table 3 (in particular, those by Bate et al. 2008, 2018;  Poindexter et al. 2008; Rojas et al. 2014) also support the shallower AD temperature profiles (steeper radius-wavelength dependence, $p > 4/3$). However, more numerous works demonstrate the radius-wavelength dependence either consistent with the thin-disk model ($p\approx4/3$, Eigenbrod et al. 2008; Mosquera et al. 2011) or shallower ($p < 4/3$, Blackburne et al. 2011;  Muñoz et al. 2011; Jiménez-Vicente et al. 2014; Motta et al. 2017) that suggests the temperature profiles steeper than the thin disk prediction, in conflict with the above arguments in favor of the shallower temperature profiles.

\begin{table} [ht]
\centering
\caption{ $r_{\tau=1}(V)$, $r_{\tau=1}(R)$ (in centimeters) and $T_{eff}$, calculated from Eqs. (\ref{equation ropt}) and  (\ref{equation teff}) for some sets of parameters
$\alpha$ and $A$, $\dot m$ is a dimensionless accretion rate, $\dot m=\dot M/\dot M_{cr}$, where $ \dot M_{cr}\approx1.4\cdot10^{17} (M_{BH}/M_\odot)$ g/s}
\label{two sets parameters}
\begin{tabular}{lcccc}
\hline
Parameters \: $\alpha$, \: $A$ \: and \: $\dot m$ \: \: \:& $T_{eff}$ $(^\circ K)$ \: & $r_{\tau=1}(R)$ (cm) \: & $r_{\tau=1}(V)$ (cm) \: & $\Delta t_{RV}$ (days) \:\\
\hline
$\alpha=0.035$; \ \  $A=115$; \ \  $\dot m$=28 & $1.794\cdot 10^{4}$ & $1.043\cdot 10^{17}$ & $9.666\cdot 10^{16}$ & 2.939\\
$\alpha=0.049$; \ \ $A=71$; \ \ \ \   $\dot m$=28  &$3.169\cdot 10^{4}$ & $6.545\cdot 10^{16}$ & $6.067\cdot 10^{16}$ & 1.845\\
$\alpha=0.080$; \ \  $A=145$; \ \  $\dot m$=28 & $3.352\cdot 10^{4}$ & $4.437\cdot 10^{16}$ & $4.113\cdot 10^{16}$ & 1.251\\
$\alpha=0.095$; \ \  $A=135$; \ \  $\dot m$=28 & $4.074\cdot 10^{4}$ & $3.625\cdot 10^{16}$ & $3.361\cdot 10^{16}$ & 1.022\\
\hline
$\alpha=0.023$; \ \  $A=90$; \ \ \ \   $\dot m$=24 & $1.727\cdot 10^{4}$ & $1.218\cdot 10^{17}$ & $1.129\cdot 10^{17}$ & 3.434\\
$\alpha=0.043$; \ \ $A=79$; \ \ \ \   $\dot m$=24  &$3.276\cdot 10^{4}$ & $5.995\cdot 10^{16}$ & $5.557\cdot 10^{16}$ & 1.689\\
$\alpha=0.063$; \ \  $A=140$; \ \  $\dot m$=24 & $3.393\cdot 10^{4}$ & $4.443\cdot 10^{16}$ & $4.118\cdot 10^{16}$ & 1.252\\
$\alpha=0.083$; \ \  $A=110$; \ \  $\dot m$=24 & $4.972\cdot 10^{4}$ & $3.130\cdot 10^{16}$ & $2.902\cdot 10^{16}$ & 0.882\\
\hline
$\alpha=0.015$; \ \ $A=75$; \ \ \ \   $\dot m$=17  &$2.070\cdot 10^{4}$ & $1.064\cdot 10^{17}$ & $9.865\cdot 10^{16}$ & 2.999\\
$\alpha=0.025$; \ \ $A=35$; \ \ \ \   $\dot m$=17  &$4.992\cdot 10^{4}$ & $5.216\cdot 10^{16}$ & $4.835\cdot 10^{16}$ & 1.471\\
$\alpha=0.035$; \ \  $A=96$; \ \ \ \ $\dot m$=17 & $3.909\cdot 10^{4}$ & $4.442\cdot 10^{16}$ & $4.117\cdot 10^{16}$ & 1.252\\
$\alpha=0.055$; \ \  $A=106$; \ \  $\dot m$=17 & $5.585\cdot 10^{4}$ & $2.768\cdot 10^{16}$ & $2.566\cdot 10^{16}$ & 0.780\\
\hline
\end{tabular}
\end{table}

\section{Effect of super-Eddington accretion on the apparent disk size}  \label{Apparent disk}   

The classic work by Shakura \& Sunyaev (1973) contains one of the earliest (if not the earliest one) analytical consideration of physical manifestations of the super-Eddington accretion mode. They have shown, in particular, that super-Eddington accretion flow can produce a powerful outflow (a wind), that leads to formation of an optically thick scattering envelope. They note that the electromagnetic spectrum of the observed radiation strongly depends on the density of the outflowing matter, which in turn is determined by the accretion rate and efficiency of the angular momentum transfer mechanism. If the mass accretion rate $\dot M$  exceeds the critical Eddington accretion rate ($ \dot M_{cr}\approx1.4\cdot10^{17} (M_{BH}/M_\odot)$ g/s) insignificantly, namely, at $\dot M/\dot M_{cr} < 4$, the disk radiation is re-emitted by the outflowing matter with virtually no change in its spectral properties. At higher accretion rates (that is, at $\dot M/\dot M_{cr} > 4$), the observed radiation is formed in the outflowing matter and the shape of its spectrum is undergoing significant changes. The outflowing matter flattens an AD temperature profile, and since it is opaque to the disk radiation due to Thompson’s scattering, it makes an apparent disk size larger. These ideas have been further developed both theoretically and in simulations in a number of more recent works, (Nishiyama et al. 2007; Abolmasov \& Shakura 2012; Sakurai et al. 2016; Cheng et al. 2019; Kitaki et al. 2021).

Shakura \& Sunyaev (1973) give the following expression for the envelope radius defined as the radius at which the optical density $\tau$ as observed along the light ray from a distant observer, reaches unity:

\begin{equation}
\label{equation ropt}
r_{\tau=1}\simeq10^7 \alpha^{-3/4}\left (\frac {10^{6\, \circ}K}{T}\right )^{3/8}
\left (\frac {10^{15}Hz}{\nu}\right )^{1/2}\dot m^{9/8}m^{3/4}.
\end{equation}
Here, dimensionless black hole mass $m=M_{BH}/M_\odot$, accretion rate $\dot m=\dot M/\dot M_{cr}$, temperature $(10^{6\,\circ}K)/T$  and frequency $(10^{15}Hz)/\nu$ are introduced,  $\alpha$ is the efficiency of the angular momentum transport, and $ \dot M_{cr}\approx1.4\cdot10^{17} (M_{BH}/M_\odot)$ g/s is a critical accretion rate, at which the full energy release in the disk is equal to the Eddington luminosity $ L_{Edd} $.

Shakura \& Sunyaev (1973) also give an expression for the effective temperature of the envelope $T_{eff}$:
\begin{equation}
\label{equation teff}
T_{eff} \simeq 2\cdot10^{10}\dot m^{-15/11}m^{-2/11}\alpha^{10/11}A^{-6/11} (^\circ K).
\end{equation}

Parameter $A$  characterizes a ratio of energy losses in the Compton processes to those in the free-free transitions, which may range from 10 to 300 (Shakura \& Sunyaev 1973).

Adopting $M_{BH}=0.88\cdot10^9 M_\odot$ for the black hole mass (Peng 2006) and $\dot m = 28$ for the SBS 1520+530 dimensionless accretion rate reported by Abolmasov \& Shakura (2012), one can obtain from expressions  (\ref{equation teff}) and (\ref{equation ropt}) the values of $r_{\tau=1}$ for a set of parameters $\alpha$ and $A$ selected from the range of their permissible values, $\alpha$$\ll$1 and $A$ varying within 10 – 300. Having calculated $r_{\tau=1}$ for two frequencies corresponding to the effective wavelengths of the $V$ and $R$ spectral bands, and supposing the layers of the envelope with just these radii are responsible for the observed reverberation signals, we can further estimate the expected time lag using a simplified expression similar to (\ref{equation tau}).

Calculations show that, for the specified BH mass and accretion rate, many different combinations of parameters $\alpha$ and $A$, selected arbitrarily from their range indicated above, provide the values of the time lag close to that obtained in the present work. Table \ref{two sets parameters} contains several examples of the time lag $\Delta t_{RV}$ calculated from Eqs. (\ref{equation ropt}) and (\ref{equation teff}) with the arbitrarily selected parameters $\alpha$ and $A$, among which one can see several "lucky hits", when the calculated values turned out to be very close to the value $< \! \Delta t_{RV} \! >$ = 1.25  days measured in our work.

A fairly large number of such "lucky hits" suggests that, apparently, there is a certain area on the plane of parameters $A$ and $\alpha$ , where the inter-band time lags $\Delta t_{RV}$  calculated from expressions (\ref{equation ropt}) and (\ref{equation teff}) are consistent with our measurements. To reveal this area that may be treated as the zone of the best coincidences, we plotted, in coordinates $\alpha$ and $A$, a map of differences between the calculated values of $\Delta t_{RV}$ and the value $< \! \Delta t_{RV} \! >$ = 1.25 days measured in our work (Fig. \ref{maps all masses}).  The absolute values of these differences are represented by a grayscale (in days) shown on the right side of the figure, with the smallest values corresponding to the lightest tone.

We see that, indeed, the values of  $\alpha$ and $A$, which ensure the closeness of the calculated values of $\Delta t_{RV}$  to that measured in our work, occupy a certain region that concentrates near very small values of $\alpha$ and extends up to very large values in $A$.

\section{Discussion}   \label{Discuss}  

Thus, we analyzed a discrepancy between the values of the inter-band time lag for filters $V$ and $R$ determined in our work for SBS 1520+530 and those expected from the classical thin disk model, and considered two possible ways to reconcile them.

One way is to assume that the observed reverberation signals arise in an AD with a temperature profile that is noticeably flatter as compared to the thin disk model. Namely, the temperature profile slope as low as $\beta \approx 0.375$ reproduces the time lag between the $V$ and $R$ light curves determined in our work, and is also consistent:

\begin{itemize}
\item[1)] with $\beta $  = 0.57 by Gaskell (2008) who proposed to determine the slope of an AD temperature profile from the spectral index of a composite spectrum in the UV/optical range;

\item[2)] with the result of Cornachione \& Morgan (2020) who concluded that discrepancy between the microlensing-based and luminosity-based AD sizes can be eliminated by a temperature profile slope $0.37 < \beta < 0.56$;

\item[3)] with the results obtained by Li et al. (2019) within their phenomenological model which invokes a strong wind from the disc that flattens its radial temperature profile;

\item[4)] finally, it is also consistent with the results of 25\% multicolor microlensing observations from different works collected in Table \ref{betta}.

\end{itemize}

Another approach to explain the discrepant estimates of the inter-band time delays is to admit that reverberation responses can arise in an optically dense scattering envelope produced from a massive matter outflow in a super-Eddington accretion mode. Due to multiple scattering, the envelope can significantly change the apparent (observed) spatial structure of the disk.

Depending on the ratio of the effective sizes of the envelope and the disk, the latter can be either partially or completely hidden under the envelope (Shakura 2018), and for sufficiently large envelopes, the apparent size and the observed structural peculiarities of an AD will be determined by the envelope rather than the disc .

According to Eqs.  (\ref{equation ropt}) and (\ref{equation teff}), the size of an envelope has a rather weak dependence on wavelength, scaling as $r_{\tau=1} \sim \lambda^{1/2}$. For a fixed BH mass, the envelope size increases with the accretion rate, and the smaller the parameter $\alpha$, the larger the envelope and the lower the effective temperature. If the opacity does not depend on wavelength, the object will demonstrate very similar radial intensity distributions at all wavelengths.

Thus, the presence of a scattering envelope in principle solves two problems, namely, it explains both the large apparent sizes of accretion disks and their weak dependence on wavelength. The above considerations about the properties of a scattering envelope and the peculiarities of its influence on the observed picture make us take a different look at the data in Table \ref{betta}. As is known, the direct product of multicolor observations of microlensing events are estimates of the AD radii $r_{\lambda}$ at different wavelengths. The values of $p$ in the third column are the results of fitting a set of radius measurements to the AD spectral profile $r_{\lambda} \sim \lambda^p$.

The corresponding values of the temperature profile slope $\beta$ in the fourth column of Table \ref{betta} were derived from the observed AD spectral profiles according to the thin disk model that implies that if the disk spectral profile is $r_{\lambda} \sim \lambda^p$, the temperature profile will have the form $T \sim r^{-1/p}$, (or $T \sim r^{-\beta}$ since we use $\beta$ for the temperature profile slope in the text above). This would be valid if we were dealing with radiation from an accretion disk. Meanwhile, for some objects in Table \ref{betta}, if not for most, the slopes of the spectral profiles $p=1/\beta$ are less than the value inherent in the thin disk model, $p=4/3$. This may suggest that at least some of them are in the super-Eddington regime and their disks are partially or completely imbedded into a scattering envelope. For such objects, attempts to determine temperature profile slopes from their spectral profiles determined from microlensing would hardly have any sense, as the AD structural peculiarities are to one degree or another hidden (or distorted) by a fairly dense scattering envelope. In other words, the radiation we detect refers to the envelope rather than to the disk for such objects.

As is noted in Sec. \ref{Apparent disk}, at moderately super-Eddington accretion mode, the     outflowing matter has an insignificant effect on the AD spectrum, therefore, addressing, if available, the UV/optical spectral index to determine the AD temperature profile seems to be reasonable.

\section{Conclusions}  \label{Conclus}  

Thus, we examined two possible ways to reproduce the value of the inter-band time delay for the $V$ and $R$ spectral bands we obtained in the present work. These are the re-emission:
\begin{itemize}
\item[1)] in an AD with the temperature profile shallower than the classic thin-disk model predicts, and
\item[2)] in an extended scattering envelope emerged due to the super-Eddington accretion mode.

\end{itemize}

Since the wind capable of producing a scattering envelope is also regarded to be the most probable driver to flatten an accretion disk temperature profile, both explanations imply a super-Eddington accretion mode, with a dimensionless accretion rate $\dot M/\dot M_{cr} < 4$ in the first case, and $\dot M/\dot M_{cr} > 4$ in the case of
the extended scattering envelope.  We used $\dot m=28$ for the dimensionless accretion rate in our calculations of the envelope radii, which would provide the inter-band time lags obtained in our paper. This value was reported by Abolmasov \& Shakura in 2012 and corrected later in their Erratum (2013). To have an additional independent estimate of the accretion mode, we addressed the Eddington ratio, $k = L_{Bol}/\eta L_{Edd}$. Here, $L_{Bol}$  is the bolometric luminosity, $ L_{Edd} $ – the Eddington luminosity, and $\eta $ is the accretion efficiency often accepted to equal 0.1. We determined the bolometric luminosity for SBS 1520+530 from the data in table 3 by Peng et al. (2006), that provides the logarithms of monochromatic continuum luminosity  $\lambda L_{\lambda}$ for a sample of 31 gravitationally lensed AGNs and 20 non-lensed AGNs, with SBS 1520+530 among them. Assuming that $L_{Bol} = 9 [\lambda F_{\lambda}] $  (Kollmeier et al. 2006), we obtain $L_{Bol}  = 39.24\cdot10^{45} $  erg/s  for the bolometric luminosity of SBS 1520+530, which provides, with $ L_{Edd} \approx 11.4 \cdot 10^{46} $ erg/s and $\eta $ = 0.1, the Eddington ratio of $ k\approx3.4$.

Returning to the issue raised at the end of Sec. \ref{Gravlensed} about the large difference in accretion rates for objects with almost the same BH masses and AD sizes ($\dot m=17$  for Q2237+0305 and $\dot m=28$ for SBS 1520+530 according to Abolmasov \& Shakura (2012)), we turned to the Eddington ratio for Q 2237+0305 too, to compare it with that for SBS 1520+530. The bolometric luminosity for Q 2237+0305 can be found, for example, in the works by Pooley et al. (2007) and Agol et al. (2009). The average between their estimates is $ L_{Bol}\approx38.7\cdot10^{45} $ erg/s that means very close luminosities and, accordingly, nearly identical Eddington ratios for the two objects, $ k\approx3.4$, that means moderately super-Eddington accretion regimes.

Thus, the problem remains unresolved: the objects with almost the same BH masses, AD sizes and Eddington ratios demonstrate, according to the approach proposed by Abolmasov \& Shakura (2012), significantly different dimensionless accretion rates. Admitting possible existence of some unaccounted physical reasons, we also cannot exclude the possibility that the objects are not so similar in their parameters, which, for SBS 1520+530, are not based on sufficiently numerous observations and therefore may be biased. Of the two available estimates of the SBS 1520+530 BH mass, we used the estimate by Peng et al. (2006) in our calculations, which is almost twice as large as that by Ding et al. (2017). As is clear from equation (\ref{equation 9.7}), using a smaller value of the BH mass would result in an even larger discrepancy between the thin disk model predictions and results of observations. While measurements by Peng et al. (2006) are restricted to the CIV line with the use of a virial normalization from Onken et al. (2004) and do not provide the uncertainty, Ding et al. (2017) apply cross-calibration with other lines and provide the uncertainty estimate of 0.4 dex. On the one hand, with such an uncertainty, the two values of the BH mass may be regarded as marginally consistent. On the other hand, simple calculations based on equation (\ref{equation 9.7}) show that the BH mass value, say, 5 times larger than that by Peng et al. (2006) would be capable of aligning the standard thin disk model with the observations. Such a large deviation between the individual estimates is quite possible, as can be seen, for example, from the scatter of the BH mass estimates for one of the most intensively investigated object, Q 2237+0305, (can be find collected in table 4 from Berdina et al. 2021).

The uncertainty of measuring the BH mass, which is a fundamental parameter related directly to the accretion processes and energy output in quasars, still remains to be large enough and thus can be one of the most important sources of the AD size discrepancy problem. For example, as is shown by Pozo Nunez et al. (2019), the size discrepancy in the Seyfert-1 galaxy Mrk509 can be explained by the underestimation of the BH mass.

In addition, some other factors may also contribute to the AD size discrepancy problem. For example, ignoring correction for the internal AGN reddening can result in a factor of 4 to 10 underestimation of the AGN luminosity, and thus, in about 2.6 underestimation of the AD size (Gaskell 2017; Gaskell et al. 2023). Contamination of the AGN or quasar light by Diffuse Continuum Emission (DCE) from the broad emission line region is also regarded as a strong source of biasing AD size measurements (Korista \& Goad 2019; Netzer 2022). Finally, in the framework of the “lamp-post” model, the discrepant AD size can be adjusted by a proper relocation of the corona (Papadakis et al. 2022).

As of now, most reverberation studies (cited above in the Introduction), note the existence of the AD size discrepancy, while a few of them, for example, Mudd et al. (2018), Homayouni et al. (2019), Yu et al, (2020), find their measurements to be consistent with the classic thin disc model by Shakura \& Sunyaev (1973).

Returning to SBS 1520+530, we can only anticipate that new BH mass measurements will shed more light on its properties. As for unaccounted physical factors, - for example, microlensing data for SBS 1520+530 should be mentioned, which are rather intricate and ambiguous, and even allow for the possibility of interpreting with a double accretion disk, (Yan et al. 2014).

\bmhead{Acknowledgements}
The authors would like to thank the colleagues from the Institute of Radio Astronomy of the NASU and Institute of Astronomy of the KhNU for the fruitful discussions of the problems raised in the present work. The authors are also grateful to the administrations of both institutes for their attention to the work and support, both financial and mental, which helped us complete our research. And the last but not the least, the authors would like to thank an anonymous reviewer for extremely detailed and instructive comments, which have hopefully improved our work.

\section*{Statements and Declarations}

\begin{itemize}
\item \textbf{Funding}
'The authors declare that no funds, grants, or other support were received during the preparation of this manuscript.'
\item \textbf{Competing interests}
'The authors declare no competing interests.'
\item \textbf{Ethics declaration}
‘Not applicable.'
\item \textbf{Author contribution}
'These authors contributed equally to this work.'
\end{itemize}

\end{document}